\documentclass[preprint,preprintnumbers,amsmath,amssymb,nofootinbib]{revtex4}
\usepackage{graphicx}
\begin{document}

\title{Entropy evolution in warm inflation from a 5D vacuum}
\author{ $^{1,2}$ Jes\'us Romero and $^{1,2}$ Mauricio
Bellini \footnote{E-mail address: mbellini@mdp.edu.ar}}

\address{$^{1}$ Departamento de F\'{\i}sica, Facultad de Ciencias Exactas y
Naturales, Universidad Nacional de Mar del Plata, Funes 3350,
(7600) Mar del Plata,
Argentina.\\
$^2$ Consejo Nacional de Investigaciones Cient\'{\i}ficas y
T\'ecnicas (CONICET). }

\begin{abstract}
Using some ideas of Modern Kaluza-Klein theory, we examine the
evolution of entropy on a 4D Friedmann-Robertson-Walker (FRW)
brane from a 5D vacuum state, which is defined on a 5D background
Riemann-flat metric. We found that entropy production is
sufficiently important during inflation: $S > 10^{90}$, for all
the initial values of temperature $T_0 < T_{GU}$.
\end{abstract}


\maketitle
\section{Introduction}

The standard picture of inflation introduced in 1981\cite{Guth}
relied on a scalar field, called the inflaton, which during
inflation was assumed to have no interaction with any other field.
The inflationary scenario postulates that the universe underwent a
phase of very rapid, accelerated expansion in its distant past.
Observations have provided strong support for the paradigm.
Standard inflationary scenario is divided into two regimes,
slow-rolling expansion and reheating phase. It is assumed that
exponential expansion places the universe in a super-cooled phase
and subsequently thereafter the universe is reheated. During the
inflationary epoch particles are being continually produced but
their density is rapidly diminished by the expansion of the
universe. In this scenario the radiation energy density $\rho_r$
becomes negligible rapidly since it scales inversely with the
fourth power of the scale factor. In such a case, a short time
reheating period terminates the inflationary period initiating the
radiation dominated epoch. Inflation in presence of non-negligible
radiation component is characterized by a non-isentropic
expansion. This can be realized in the warm inflation
scenarios\cite{berera}.

What distinguishes warm inflation is that this particle production
is sufficiently strong compared with the effects of the expansion,
so that it is possible to produce a non-negligible particle
density. Warm inflation occurs when the particle number is large
enough to influence the classical inflaton field and produce
density fluctuations\cite{bel1}. Two important facts in warm
inflation are that it can provide a graceful exit to the standard
inflation dynamics, and can give a physical justification of the
slow-rolling conditions, which are incorporated ad hoc in models
of standard inflation.

In this paper we examine the evolution of entropy on a 4D brane
from a 5D vacuum state, which is defined on a 5D background
Riemann-flat metric, using ideas of Modern Kaluza-Klein theory.
This theory allow the fifth coordinate to play an important
physical role. In the framework of the Induced Matter (or
Space-Time-Matter) theory\cite{wbook}, all classical physical
quantities, such as matter density and pressure, are susceptible
of a geometrical interpretation. The mathematical basis of it
relies in the Campbell's theorem\cite{campbell}, which ensures an
embedding of 4D general relativity with sources in a 5D theory
whose field equations are apparently empty. That is, the Einstein
equations\footnote{In this paper we use capital Latin letters that
run from $0$ to $4$ and Greek letters that run from $0$ to $3$. }
$G_{\alpha\beta}=-8\pi G\,T_{\alpha\beta}$ (we use natural units:
$c=\hbar=1$), are embedded perfectly in the Ricci-flat equations
$R_{AB}=0$. Other version of 5D gravity, which is mathematically
similar, is the membrane theory, in which gravity propagates
freely on the 5D bulk and the interactions of particles are
confined to a 4D hypersurface called "brane"\cite{rs}. Both
versions of 5D general relativity are in agreement with
observations.

\section{5D vacuum and field dynamics}

We consider the recently introduced 5D Riemann-flat
metric\cite{MET}
\begin{equation}\label{a1}
dS^{2}=\psi^{2}\,\frac{\Lambda(t)}{3}\,dt^{2}-\psi ^{2} \,e^{2\int
\sqrt{\Lambda(t)/3}\,dt}dr^{2}-d\psi ^{2},
\end{equation}
where $\Lambda(t)$ is a time-dependent function, which can be
interpreted as a decaying cosmological parameter under certain
circumstances\cite{rus}. Furthermore, $dr^{2}=dx^2+dy^2+dz^2$ is
the 3D Euclidean metric, $t$ is the cosmic time and $\psi$ is the
space-like noncompact extra dimension. Since the metric (\ref{a1})
is Riemann-flat (and therefore Ricci-flat), hence it is suitable
to describe a 5D vacuum vacuum ($G_{AB}=0$) in the framework of
Space-Time-Matter (STM) theory of gravity\cite{STM}. With this aim
we shall consider the 5D action
\begin{equation}\label{act}
^{(5)}I = {\Large\int} d^4 x \  d\psi \sqrt{\left|\frac{^{(5)}
 g}{^{(5)} g_0}\right|} \left(
\frac{^{(5)} R}{16\pi G}+ \frac{1}{2}g^{AB} \varphi_{,A}
\varphi_{,B} \right),
\end{equation}
where $^{(5)}g$ is the determinant of the covariant metric tensor
$g_{AB}$:
\begin{equation}\label{ricci5}
^{(5)}g= \psi^8 \,\left(\frac{\Lambda}{3}\right) \, e^{6\int
\sqrt{\frac{\Lambda}{3}} dt},
\end{equation}
and $^{(5)} g_0=\psi^8_0 \left({\Lambda_0\over3}\right)$ is a
constant to make dimensionless the expression $\left|^{(5)} g
/^{(5)} g_0\right|$.

In order to describe a 5D vacuum on (\ref{a1}), we shall consider
$\varphi$ as a massless test classical scalar field, which is
minimally coupled to gravity. For this reason, the Lagrangian
related to $\varphi$ is purely kinetic and free of any interaction
on (\ref{a1}). From the mathematical point of view, the second
term in the action (\ref{act}) is constructed by using monogenic
fields $\varphi$, which have null D\'{}Alambertian on the 5D
Riemann-flat metric (\ref{a1})\cite{almeida}.

The equation of motion for the field $\varphi$ is
\begin{equation}
\ddot\varphi + \left[3\sqrt{\frac{\Lambda}{3}} -
\frac{\dot\Lambda}{2\Lambda}\right] \dot\varphi -
\frac{\Lambda}{3} e^{-2\int \sqrt{\frac{\Lambda}{3}} dt}
\nabla^2\varphi  -  \frac{\Lambda}{3}\left[4\psi
\frac{\partial\varphi}{\partial\psi} + \psi^2
\frac{\partial^2\varphi}{\partial\psi^2}\right] =0.\label{dis}
\end{equation}
Furthermore, since $\varphi$ evolves on a Riemann-flat spacetime
and complies with the Einstein equations $G_{AB}=-8\pi G\,T_{AB}$,
($G$ is the gravitational constant), the energy-momentum tensor
$T_{AB}$ must be null on (\ref{a1}).

In the next sections we shall study the interpretation for the
dynamics of the scalar field $\varphi$ on an effective 4D FRW
brane.

\section{Dissipative dynamics of $\varphi$ on a FRW brane}

We shall assume that the 5D spacetime (\ref{a1}) is dynamically
foliated on the fifth coordinate: $\psi=\psi(t)$, such that the
effective 4D hypersurface is
\begin{eqnarray}
&& \left.dS^2\right|_{\psi=\psi(t)}=
\left.d\sigma^2\right|_{\psi=\psi(t)} -\dot\psi^2(t) dt^2
\nonumber \\
&& \equiv \left[\psi^2(t)\frac{\Lambda(t)}{3} -
\dot\psi^2(t)\right]dt^2 - \psi^2(t) \, e^{2\int
\sqrt{\frac{\Lambda(t)}{3}} dt} \, dr^2, \label{u}
\end{eqnarray}
with the condition $\psi^2(t) {\Lambda(t)\over 3} - \dot\psi^2(t)
>0$, such that $g_{AB} U^A U^B =1$ [$U^A = {dX^A\over dS}$ are the
components of the penta-velocity]. We can define
$\rho=\rho_0+\Delta\rho$ and $p=p_0+\Delta p$, the energy density
and the pressure on $\left.d\sigma^2\right|_{\psi=\psi(t)}$ in
(\ref{u}).

\subsection{Entropy evolution on the $d\sigma^2$}

We shall consider that $\rho_0(t)$ and $p_0(t)$ are the energy
density and the pressure on the effective 4D hypersurface
$\left.dS^2\right|_{\psi=\psi(t)}$ in (\ref{u}) and the system
describes an adiabatic evolution on (\ref{u}), so that (from the
thermodynamical point of view) it can be considered as a closed
system
\begin{equation}
\frac{d}{dt}\left[\rho_0\,a^3(t)\right] + p_0\,
\frac{d}{dt}\left[a^3(t)\right]=0.
\end{equation}

Since the universe describes a de Sitter expansion during the
inflationary epoch, it is interesting to study the evolution of
$\varphi$ on the hypersurface
$\left.d\sigma^2\right|_{\psi=\psi(t)}= g_{\mu\nu}\left(t,\vec
r,\psi(t)\right) \, dx^{\mu} dx^{\nu} \neq
\left.dS^2\right|_{\psi=\psi(t)}$, which can be considered as a
brane on (\ref{u}). It is very important to remark that the
equivalence principle is broken on
$\left.d\sigma^2\right|_{\psi=\psi(t)}$: $g_{AB} U^A U^B \neq 1$.
Hence, $\Delta\rho$ and $\Delta p$ comply with the following
equation on $\left.d\sigma^2\right|_{\psi=\psi(t)}$:
\begin{equation}\label{entropia}
\Delta\dot\rho+3 H \,\gamma \,\Delta\rho
=\frac{T}{a^3(t)}\dot{\cal S}.
\end{equation}
where $\gamma = 1+{\Delta p\over \Delta\rho}$,
$a(t)=\psi(t)\,e^{\int \sqrt{{\Lambda\over 3}}\,dt}$ is the scale
factor on $\left.d\sigma^2\right|_{\psi=\psi(t)}$, $H(t)=\dot a/a$
is the Hubble parameter, $T$ is the background temperature of the
thermal bath, and ${\cal S}$ is its entropy. A very interesting
case, of particular interest is ${\Delta p\over \Delta\rho}=1/3$.
In this particular case we can identify $\Delta\rho$ and $\Delta
p$ with the radiation energy density and its pressure: $\Delta\rho
\equiv \rho_r$ and $\Delta p \equiv p_r$. In this case $\gamma=
4/3$ and the system radiates
\begin{equation}\label{en}
\dot\rho_r+4 H\,\rho_r =\delta,
\end{equation}
where the interaction $\delta\equiv\frac{T}{a^3(t)} \dot{\cal S}
> 0$ is related with the variation of entropy\cite{rb}. Notice
that the entropy increases with time, so that both sides in
(\ref{en}) are positives. The interesting is that eq. (\ref{en})
is the same to those of warm inflation, for the evolution of
radiation energy density.

\subsection{Einstein equations on $d\sigma^2$}

To understand better the equations (\ref{entropia}) and
(\ref{en}), we can start to write the Energy Momentum tensor for a
perfect fluid on the 5D Riemann-flat metric (\ref{a1})
\begin{equation}
T^A_{\,\,B}=(P+\rho) U^A U_B - P\, g^A_{\,\,B},
\end{equation}
such that $U^A U_A=1$ and $dS^2=\left(\psi^2 \Lambda/3\right)
\,ds^2 -d\psi^2$. Furthermore, we take
\begin{equation}
\left(\frac{ds}{dS}\right)^2 = \frac{3}{\Lambda \psi^2} \left[ 1 +
U^4 U_4\right].
\end{equation}
In our case, we are choosing $U^4 = \dot\psi U^0$ [see the metric
(\ref{u})], so that if we take a co-moving frame $U^i=u^i=0$, we
obtain the following expression:
\begin{equation}
u^0 u_0 = \frac{1}{\left[1+\dot\psi^2 U^0 U_0\right]} \leq 1,
\end{equation}
where the equality holds when $\dot\psi=0$.

On the brane $d\sigma^2$, the energy momentum tensor has the form
\begin{equation}
\left. T^{\alpha}_{\,\,\beta}\right|_{\psi(t)} = \left.\left(P+\rho\right) \left(\frac{3}{\psi^2\Lambda}\right)
u^{\alpha} u_{\beta} - P \delta^{\alpha}_{\,\,\beta} + \left(P+\rho\right)  \left[g^{\alpha}_{\,\,\beta}
-\left(\frac{3}{\psi^2\Lambda}\right)
u^{\alpha} u_{\beta}\right]\right|_{\psi(t)}.
\end{equation}
In particular, if we choose $\psi(t)=\sqrt{{3\over\Lambda}}$, we
obtain
\begin{eqnarray}
&& \left. T^{0}_{\,\,0}\right|_{\psi(t)=\sqrt{{3\over\Lambda}}} = \rho, \\
&& \left. T^{x}_{\,\,x}\right|_{\psi(t)=\sqrt{{3\over\Lambda}}}=\left. T^{y}_{\,\,y}\right|_{\psi(t)=\sqrt{{3\over\Lambda}}}=
\left. T^{z}_{\,\,z}\right|_{\psi(t)=\sqrt{{3\over\Lambda}}}= -  P,
\end{eqnarray}
where $P=-\rho$, and
\begin{eqnarray}
&& \rho_{\varphi} = \rho\,u^{0} u_{0},\\
&& \rho_{r} =\rho\,\left[1-u^{0} u_{0}\right].
\end{eqnarray}
Here, the total energy density on the brane $d\sigma^2$ is given
by matter and radiation energy densities
($\rho_{\varphi},\rho_r$): $\rho = \rho_{\varphi} + \rho_r$ and
the total pressure $P$ is given by matter $P_{\varphi}$ and
radiation $P_r$, such that $P_r=\rho_r/3$:
$P=P_{\varphi}+\rho_r/3$.

The Friedmann equations on the metric $d\sigma^2$ are
\begin{eqnarray}
&& 3 H^2 = -8\pi G\, \left[\rho_{\varphi} + \rho_r\right], \\
&& 3 H^2 + 2\dot H = 8\pi G\, \left[P_{\varphi}+\rho_r/3\right],
\end{eqnarray}
which are exactly the same as those of warm inflation.

\section{An example}

Now we consider the foliation where $\psi^2(t)=3/\Lambda(t)$. In
this case the effective 4D metric (\ref{u}) is
\begin{equation}\label{close}
\left.dS^2\right|_{\psi(t)=\sqrt{{3\over\Lambda(t)}}} =
\left.d\sigma^2\right|_{\psi(t)=\sqrt{{3\over\Lambda(t)}}}-
\frac{3}{\Lambda(t)}
\left(\frac{\dot\Lambda(t)}{2\Lambda(t)}\right)^2\, dt^2,
\end{equation}
where
\begin{equation}\label{uuu}
\left.d\sigma^2\right|_{\psi(t)=\sqrt{{3\over\Lambda(t)}}}=dt^2 -
\frac{3}{\Lambda(t)} \, e^{2\int \sqrt{\frac{\Lambda(t)}{3}} dt}
\, dr^2\equiv  dt^2 - a^2(t) \, dr^2.
\end{equation}
From the thermodynamical point of view, the hypersurface described
by the brane
$\left.d\sigma^2\right|_{\psi(t)=\sqrt{{3\over\Lambda(t)}}}$ can
be considered as an open system with respect to the closed one
$\left.dS^2\right|_{\psi(t)=\sqrt{{3\over\Lambda(t)}}}$ in
(\ref{close}). We shall consider that (\ref{uuu}) is our physical
spacetime.

The equation of motion for the field $\varphi(t,\vec r)$ on the
metric (\ref{uuu}) [induced from the 5D vacuum defined by the
action (\ref{u}), on the metric (\ref{a1})], is
\begin{equation}
\ddot\varphi + 3 \frac{\dot a}{a} \dot\varphi - \frac{1}{a^2} \,
\nabla^2_r\varphi - \left.\frac{\Lambda(t)}{3}\left[ 4\psi
\frac{\partial\varphi}{\partial\psi} +
\psi^2\frac{\partial^2\varphi}{\partial\psi^2}\right]
\right|_{\psi(t)=\sqrt{\frac{3}{\Lambda(t)}}}=-\frac{\delta}{\dot\varphi},\label{y}
\end{equation}
where the interaction on the right-hand side of (\ref{y}) has its
origin in the temporal dependence of the fifth coordinate on the
brane (\ref{uuu}).

\begin{figure*}
\includegraphics[totalheight=8.5cm]{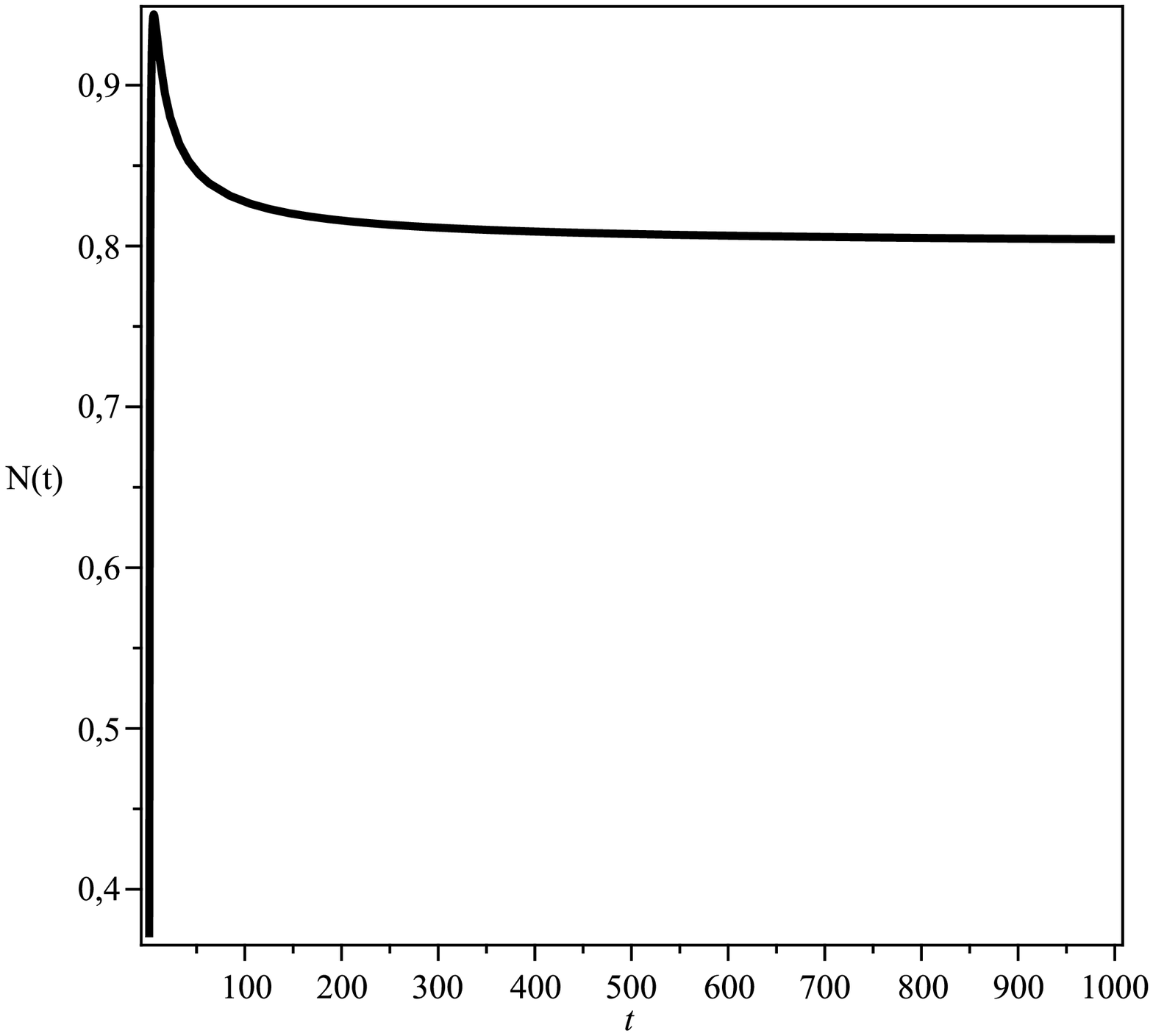} \caption{{\small  Evolution
of $\Gamma/(3H)=-\dot\Lambda/[3 \Lambda\, H]$ as a function of
cosmic time $t$ [expressed in Planckian times $G^{-1/2}$]. Notice
that dissipation is of the order of expansion during all the
inflationary phase.} \label{fig1}}
\end{figure*}

In order to study the evolution of entropy ${\cal S}$ during warm
inflation, we shall consider the case where
\begin{equation}\label{gama}
\gamma=\frac{4}{3}-\frac{1}{3}\left({t-t_0}\over t\right),
\end{equation}
where $t\geq t_0$, $t_0$ being the time when inflation starts.
Notice that at the beginning of inflation (i.e., for $t \simeq
t_0$ the parameter $\gamma \simeq 4/3$ and $\Delta P/\Delta \rho
\simeq 1/3$. However, after a few Planckian times one recovers
$\Delta P/\Delta \rho \simeq 0$, so that $\gamma \simeq 1$, for a
matter dominated universe. In other words, the evolution of $\rho$
is governed by the equation
\begin{equation}\label{a}
\dot\rho+ \gamma (t) H \,\rho =\delta ,
\end{equation}\label{b}
where  $\gamma(t)$ is given by eq. (\ref{gama}). We consider
$a(t)=a_0\,\left(t/t_0\right)^{[n(t)+1]}$, so that
$n(t)=\left({t_0\over t}\right)$. In this case the Hubble
parameter is
\begin{equation}\label{c}
H= \dot n \,{\rm ln}(t/t_0)+ \left({n+1\over t}\right)=
\frac{t_0}{t^2} \left[1-{\rm ln}(t/t_0)\right] +\frac{1}{t}.
\end{equation}
In the figure (\ref{fig1}) we have plotted the evolution of
$\Gamma/(3H)$, as a function of time, such that
$\Gamma(t)=\delta/\dot\varphi^2$. Notice that dissipation $\Gamma$
is of the order of the expansion $3H$ during all the inflationary
stage, so that the production of entropy remains high in this
epoch.

The evolution of entropy is given by
\begin{equation}\label{d}
\dot S= {a^4 \over 2 \pi T_0 a_0} {{\dot H}^2 \over
H\,G},\end{equation}
so that
\begin{equation}\label{e}
S(t)=-{1\over {2 \pi T_0 a_0\,G}}\int_{t_0}^t a^4(\tau)
\left(\frac{\dot{H}^2}{H}\right)\,\,d\tau.
\end{equation}

In the next table we endorse the entropy produced during warm
inflation for  $ k = 1.5$ and $p = 1$, taking into account
different values of initial temperature $T_0$

\begin{center}
\begin{tabular}{|c|c|c|c|}
  \hline
   k = 1.5 ; p = 1

 & $S \sim 10^{90}$ & $S \sim 10^{94}$ & $S \sim 10^{98}$ \\
  \hline
  $T_0 = 10^{-19}\,\,G^{-1/2}$ & $t_f = 10^9\,\,G^{1/2}$ & $t_f = 10^{9.5}\,\,G^{1/2}$ & $t_f = 10^{10}\,\,G^{1/2}$ \\
  $T_0 = 10^{-15}\,\,G^{-1/2}$ & $t_f = 10^{9.5}\,\,G^{1/2}$  & $t_f = 10^{10}\,\,G^{1/2}$ & $t_f = 10^{10.5}\,\,G^{1/2}$ \\
  $T_0 = 10^{-11}\,\,G^{-1/2}$ & $t_f = 10^{10}\,\,G^{1/2}$ & $t_f = 10^{10.5}\,\,G^{1/2}$ & $t_f = 10^{11}\,\,G^{1/2}$ \\
  $T_0 = 10^{-7}\,\,G^{-1/2}$ & $t_t = 10^{10.5}\,\,G^{1/2}$ & $t_f = 10^{11}\,\,G^{1/2}$ & $t_f = 10^{11.5}\,\,G^{1/2}$
  \\
  \hline
\end{tabular}
\end{center}
where we are using natural units $\hbar=c=1$, so that the
Planckian mass is given by $M_p =G^{-1/2}$ and $G$ is the
gravitational constant. The interesting here is that models with
smaller initial temperature generate more entropy in shortest
times.

\section{Final Comments}

In this letter we have studied the evolution of entropy on a 4D
FRW metric, which can be considered as a brane on a 5D
Riemann-flat metric on which we define a vacuum. The generation of
entropy is a consequence of dissipative dynamics of the inflaton
field, which can be induced from a 5D vacuum state. Furthermore,
the equivalence principle is broken on the FRW brane on which the
"extra force" that acts on the inflaton field must be interpreted
as non-conservative.

In the example here studied we have examined different
possibilities of evolution with different initial conditions for
warm inflation. Our results show how the rate of production of
entropy of the universe increases during inflation when $T_0$ is
smallest. In this sense, models of warm inflation with smallest
initial temperature are favored because they produce more entropy
in shortest times. Anyway, we found that for all $T_0$ which are
below $T_{GU}\simeq 10^{15}\,\,G^{-1/2}$, warm inflation produces
a sufficiently high amount of entropy to assure sufficiently
flatness of the universe: $S
> 10^{90}$.

\begin{acknowledgments}
The authors acknowledge CONICET and UNMdP (Argentina) for
financial support.
\end{acknowledgments}

\newpage

\end{document}